\documentclass[aps,twocolumn]{revtex4-1}
\usepackage{amsfonts}
\usepackage{amssymb}
\usepackage{amsmath}
\usepackage{graphicx}
\usepackage{epsfig}
\usepackage{color}

\begin{document}

\title{Steady helix states in a resonant XXZ Heisenberg model with
Dzyaloshinskii-Moriya interaction}
\author{E. S. Ma}
\author{K. L. Zhang}
\author{Z. Song}
\email{songtc@nankai.edu.cn}
\affiliation{School of Physics, Nankai University, Tianjin 300071, China}
\begin{abstract}
We systematically investigate possible helix states in XXZ Heisenberg model
with Dzyaloshinskii-Moriya (DM) interaction. Exact solutions show that a set
of precession helix states can be constructed by deliberate superposition of
degenerate eigenstates of the Hamiltonian under the resonant condition. When
a non-Hermitian balance boundary term is imposed as a quenching action, the
quench dynamics shows that a steady helix state emerges from some easily
prepared initial states, including saturate and maximally mixed
ferromagnetic states, according to the analysis of perturbation method. The
corresponding dynamics for near resonant cases is also investigated
numerically, indicating the robustness of the scheme. Our findings highlight
the cooperation of non-Hermiticity and the DM interaction in quantum spin
system, suggesting a way for preparing steady helix state in non-Hermitian
quantum spin system.
\end{abstract}

\maketitle




\section{Introduction}

The quantum Heisenberg model, as a simple model of interacting spins, takes
an important role in physics. It not only captures the properties of many
magnetic materials, but also provides a tractable theoretical example for
understanding fundamental concepts in physics. Although the one-dimensional
Heisenberg chain is a old topic, quantum dynamics of the system is still an
active frontier of research, especially after the quantum simulator is
realized in experiment \cite%
{zhang2017observation,bernien2017probing,barends2015digital,davis2020protecting,signoles2021glassy,trotzky2008time,gross2017quantum}%
. Recently, the discovery of highly excited many-body eigenstates of the
Heisenberg model, referred to as Bethe phantom states, has received much
attention from both theoretical \cite%
{popkov2016obtaining,popkov2017solution,popkov2020exact,popkov2021phantom}
and experimental approaches \cite%
{jepsen2020spin,jepsen2021transverse,hild2014far,jepsen2022long}.

In this work, we investigate possible helix states in XXZ Heisenberg model
under two considerations. One corresponds to the introduction of
Dzyaloshinskii-Moriya (DM) interaction. The DM interaction is an
antisymmetric exchange interaction that appears in inversion asymmetric
structures and favors perpendicular alignment of neighboring spins in a
magnetic material \cite%
{dzyaloshinsky1958thermodynamic,bode2007chiral,roessler2006spontaneous}. The
other is the imposed non-Hermitian balance boundary condition, which takes
the role of source and drain of spin flip. Under a resonant condition on the
DM and anisotropic terms, the modified Heisenberg model obeys the SU(2)
symmetry, and then possesses a set of degenerate eigenstates. It allows the
existence of spin helix state as exact solution obtained by deliberate
superposition of these degenerate eigenstates. We are interested in the
dynamic preparation of the spin helix state. Based on the analysis of
perturbation method, it is shown that a steady helix state emerges from some
easily prepared initial states, including saturate and maximally mixed
ferromagnetic states, when a non-Hermitian balance boundary is imposed as a
quenching action. For near resonant cases, the corresponding dynamics is
also investigated numerically and the results indicate that the scheme works
well at certain time window. It relates to an exclusive concept in a
non-Hermitian system, exceptional point (EP), which has no counterpart in a
Hermitian system. The EP in a non-Hermitian system occurs when eigenstates
coalesce \cite{bender2007making, moiseyev2011non, krasnok2019anomalies}, and
usually associates with the non-Hermitian phase transition \cite%
{feng2013experimental, gupta2019parity}. In a parity-time ($\mathcal{PT}$)
symmetric non-Hermitian coupled system, the $\mathcal{PT}$ symmetry of
eigenstates spontaneously breaks at the EP \cite%
{guo2009observation,ruter2010observation,peng2014parity,feng2014single,hodaei2014parity,feng2017non}%
, which determines the exact $\mathcal{PT}$-symmetric phase and the broken $%
\mathcal{PT}$-symmetric phase in this system.

In this work, we will impose a pair of balance non-Hermitian impurities \cite%
{zhang2020resonant,zhang2020dynamic} to the ends of the spin chain, as
non-Hermitian boundary condition. The corresponding dynamics is also
investigated analytically and numerically. The approximate solutions for the
quantum spin chain with finite length provide valuable insights for the
description of the non-equilibrium dynamics. Our findings highlight the
cooperation of non-Hermiticity and the DM interaction in quantum spin
system, suggesting a way for preparing steady helix state in non-Hermitian
quantum spin system.

The rest of this paper is organized as follows: In Sec.~\ref{Model
Hamiltonian}, we introduce the model Hamiltonian and the corresponding SU(2)
symmetry. With these preparations, in Sec.~\ref{Two types of helix states}
we demonstrate that two types of helix states can be constructed by a set of
degenerate eigenstates. Based on these results, the dynamic generation of
spin helix state are proposed in Sec.~\ref{Dynamic generation of helix state}
by means of three kinds of imposed fields. Sec.~\ref{Summary} concludes this
paper.

\section{Model Hamiltonian and symmetries}

\label{Model Hamiltonian}

We begin this section by introducing a general Hamiltonian 
\begin{equation}
H=H_{0}+H_{\mathrm{I}}  \label{H spin}
\end{equation}%
where $H_{0}$\ and $H_{\mathrm{I}}$\ describe quantum spin Heisenberg chain
with DM interaction and external interaction respectively%
\begin{align}
H_{0}=-\sum_{j=1}^{N-1}\left(
J_{x}s_{j}^{x}s_{j+1}^{x}+J_{y}s_{j}^{y}s_{j+1}^{y}+J_{z}s_{j}^{z}s_{j+1}^{z}\right) &
\notag \\
+i\frac{D}{2}\sum_{j=1}^{N-1}\left(
s_{j}^{+}s_{j+1}^{-}-s_{j}^{-}s_{j+1}^{+}\right) ,H_{\mathrm{I}}=\sum_{j}^{N}%
\mathbf{B}_{j}\cdot \mathbf{s}_{j}.&
\end{align}%
Here $\mathbf{s}_{j}=\left( s_{j}^{x},s_{j}^{y},s_{j}^{z}\right) $\ is the
spin-$1/2$ operator, and $\mathbf{B}_{j}$ is on-site magnetic field,
inducing Hermitian or non-Hermitian impurity. In this work, we only focus on
the case with $J_{x}=J_{y}$, and by taking $\left( J_{x}\right) ^{2}+D^{2}=1$
and $\Delta =J_{z}$ for the sake of simplicity, we rewrite $H_{0}$\ as the
form 
\begin{equation}
H_{0}=-\sum_{j=1}^{N-1}\left( \frac{e^{-ik_{0}}}{2}s_{j}^{+}s_{j+1}^{-}+%
\frac{e^{ik_{0}}}{2}s_{j}^{-}s_{j+1}^{+}+\Delta s_{j}^{z}s_{j+1}^{z}\right) ,
\end{equation}%
where $k_{0}=\arctan (D/J_{x})$ is a crucial factor for helix state arising
from $D$. For arbitrary $\Delta $, we always have%
\begin{equation}
\left[ s^{z},H_{0}\right] =0,
\end{equation}%
with $s^{z}=\sum_{j=1}^{N}s_{j}^{z}$. Importantly, for the resonant case
with $\Delta =1$ defining 
\begin{equation}
s_{k_{0}}^{+}=\left( s_{k_{0}}^{-}\right) ^{\dag }=\underset{j=1}{\overset{N}%
{\sum }}e^{ik_{0}j}s_{j}^{+},
\end{equation}%
we have

\begin{equation}
\left[ s_{k_{0}}^{\pm },H_{0}\right] =0,
\end{equation}%
which is not a surprising result since $s_{k_{0}}^{\pm }$\ and $s^{z}$\
satisfy the Lie algebra commutation relations

\begin{equation}
\left[ s_{k_{0}}^{+},s_{k_{0}}^{-}\right] =2s^{z},\left[ s^{z},s_{k_{0}}^{%
\pm }\right] =\pm s_{k_{0}}^{\pm }.
\end{equation}%
It seems a little trivial but is helpful for the following processing in the
presence of impurity term $H_{\mathrm{I}}$.

\section{Two types of helix states}

\label{Two types of helix states}

In this section, we will introduce two types of helix states based on the
eigenstates of $H_{0}$ with $\Delta =1$. We start by the ferromagnetic
eigenstate of $H_{0}$

\begin{equation}
\left\vert \psi _{0}\right\rangle =\left\vert \Downarrow \right\rangle =%
\overset{N}{\underset{j=1}{\prod }}\left\vert \downarrow \right\rangle _{j},
\end{equation}%
satisfying the equation $H_{0}\left\vert \psi _{0}\right\rangle
=-(N-1)/4\left\vert \psi _{0}\right\rangle $, with $s_{j}^{z}\left\vert
\downarrow \right\rangle _{j}=-1/2\left\vert \downarrow \right\rangle _{j}$.
Based on the symmetry of $H_{0}$\ mentioned above, a set of eigenstates $%
\left\{ \left\vert \psi _{n}\right\rangle ,n\in \left[ 1,N\right] \right\} $%
\ can be constructed as\ 

\begin{equation}
\left\vert \psi _{n}\right\rangle =\frac{1}{\Omega _{n}}\left(
s_{k_{0}}^{+}\right) ^{n}\left\vert \Downarrow \right\rangle ,
\end{equation}%
where the normalization factor $\Omega _{n}=\left( n!\right) \sqrt{C_{N}^{n}}
$. Obviously, we have $\left\vert \psi _{N}\right\rangle $ $=e^{ik_{0}\left(
1+N\right) N/2}\left\vert \Uparrow \right\rangle $ $=e^{ik_{0}\left(
1+N\right) N/2}\prod\nolimits_{j=1}^{N}\left\vert \uparrow \right\rangle
_{j} $. We introduce a local vector $\mathbf{h}_{l}=\left(
h_{l}^{x},h_{l}^{y},h_{l}^{z}\right) $ with $h_{l}^{\alpha }=\left\langle
\psi \right\vert s_{l}^{\alpha }\left\vert \psi \right\rangle $ ($\alpha
=x,y,z$) to characterize the helicity of a given state $\left\vert \psi
\right\rangle $.\ 

For eigenstates $\left\vert \psi _{n}\right\rangle $, straightforward
derivation of $h_{l}^{\alpha }(n)=\left\langle \psi _{n}\right\vert
s_{l}^{\alpha }\left\vert \psi _{n}\right\rangle $\ show that 
\begin{equation}
h_{l}^{x}(n)=h_{l}^{y}(n)=0,h_{l}^{z}(n)=\frac{n}{N}-\frac{1}{2},
\end{equation}%
which is uniform, indicating that $\left\vert \psi _{n}\right\rangle $\ is
not a helix state. Nevertheless, in the following we will show that their
superposition can be helix states. And these states can be classified as two
types of helix states: precession and entanglement helix states.

\subsection{Precession helix state}

\begin{figure*}[tbh]
\centering
\includegraphics[width=1\textwidth]{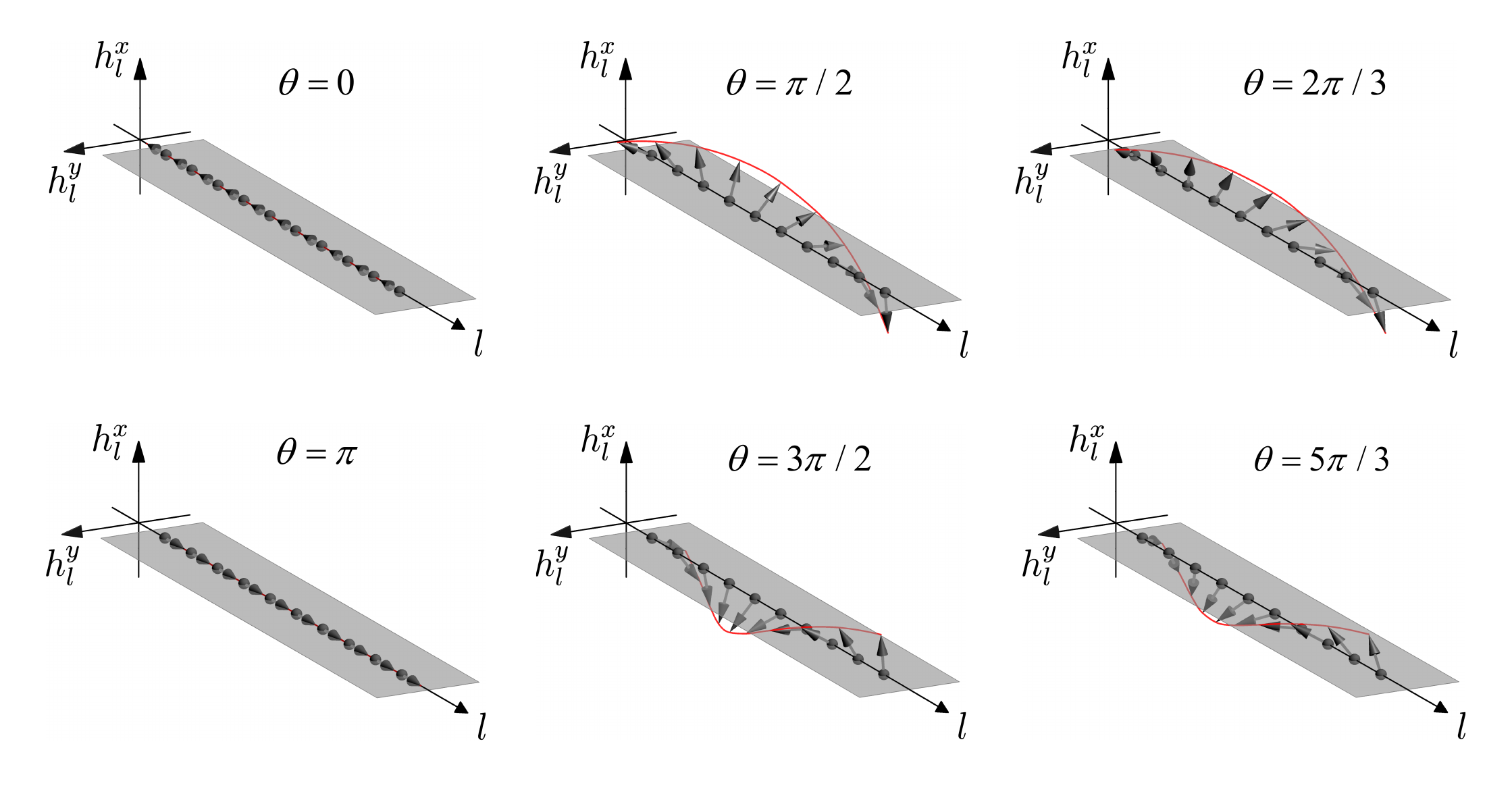}  
\caption{Plots of helix vector from Eq. (\protect\ref{state}) for several
representative values of $\protect\theta $, with parameters $k_{0}=\arctan
(0.5)$ and $N=10$. For $\protect\theta =0$ and $\protect\pi $, all spins
align in the $z$-direction. For $\protect\theta =\protect\pi /2$ and $3 
\protect\pi /2$, all spins lie in the $xy$-plane.}
\label{fig1}
\end{figure*}

We consider a superposition eigenstates in the form\ 
\begin{equation}
\left\vert \phi (\theta )\right\rangle =\sum_{n}d_{n}\left\vert \psi
_{n}\right\rangle ,  \label{state}
\end{equation}%
where%
\begin{equation}
d_{n}=\sqrt{C_{N}^{n}}\left( -i\right) ^{n}\sin ^{n}\left( \theta /2\right)
\cos ^{\left( N-n\right) }\left( \theta /2\right) .
\end{equation}%
The corresponding helix vector is 
\begin{equation}
\mathbf{h}_{l}=\frac{1}{2}[\sin \theta \sin \left( k_{0}l\right) ,\sin
\theta \cos \left( k_{0}l\right) ,-\cos \theta ],
\end{equation}%
which indicates that $\left\vert \phi (\theta )\right\rangle $\ is a helix
state for nonzero $\sin \theta $. Here $\theta $ is an arbitrary angle and
determines the profile of the state. This can be obtained easy when we
express it in the form.

\begin{equation}
\left\vert \phi (\theta )\right\rangle =\overset{N}{\underset{j=1}{\prod }}%
\left( -ie^{ik_{0}j}\sin \left( \theta /2\right) \left\vert \uparrow
\right\rangle _{j}+\cos \left( \theta /2\right) \left\vert \downarrow
\right\rangle _{j}\right) .  \label{phi}
\end{equation}%
It represents a tensor product of the precession states of all spins, which
is a unentangled state. It accords with the result $\left\vert \mathbf{h}%
_{l}\right\vert ^{2}=1/4$. Plots of $\mathbf{h}_{l}$ for several typical
cases are presented in Fig. \ref{fig1}. 

In addition, one can express state $\left\vert \phi (\theta )\right\rangle $%
\ in the form $\left\vert \phi (\theta )\right\rangle =R_{\theta
}^{+}\left\vert \Downarrow \right\rangle $, where the operator is

\begin{equation}
R_{\theta }^{\pm }=\sum_{n}\frac{d_{n}}{\Omega _{n}}\left( s_{k_{0}}^{\pm
}\right) ^{n},
\end{equation}%
satisfying%
\begin{equation}
\left[ R_{\theta }^{\pm },H_{0}\right] =0.
\end{equation}%
We note that%
\begin{equation}
\left( R_{\theta }^{+}\right) ^{m}\left\vert \phi (\theta )\right\rangle
\propto \left\vert \phi (\theta ^{\prime })\right\rangle ,
\end{equation}%
with $\tan \left( \theta ^{\prime }/2\right) =\left( m+1\right) \tan \left(
\theta /2\right) $, which indicates that the action of operator $\left(
R_{\theta }^{+}\right) ^{m}$ is a shift of the angle $\theta \longrightarrow
\theta ^{\prime }$, referred to as angle shift operator.

\subsection{Entanglement helix state}

Here is an example for entanglement helix state. We construct a state by a
simple superposition%
\begin{equation}
\left\vert \psi _{\mathrm{E}}\right\rangle =\frac{1}{\sqrt{N^{2}+N}}\underset%
{j}{\overset{N}{\sum }}\left( e^{ik_{0}j}s_{j}^{+}+1\right) \left\vert
\Downarrow \right\rangle .
\end{equation}%
The corresponding helix vector is 
\begin{equation}
\mathbf{h}_{l}=\frac{1}{2\left( N+1\right) }[2\cos \left( k_{0}l\right)
,-2\sin \left( k_{0}l\right) ,\frac{2}{N}-N-1],
\end{equation}%
which indicates that $\left\vert \psi _{\mathrm{E}}\right\rangle $\ is a
weak helix state for finite $N$. In addition, we note that%
\begin{equation}
\left\vert \mathbf{h}_{l}\right\vert ^{2}=\frac{4N^{2}+\left(
N^{2}+N-2\right) ^{2}}{4N^{2}\left( N+1\right) ^{2}},
\end{equation}%
and$\ \left\vert \mathbf{h}_{l}\right\vert ^{2}<1/4$ for finite $N$. It
indicates that $\left\vert \psi _{\mathrm{E}}\right\rangle $\ cannot be
written as a tensor product, in the form of $\left\vert \phi (\theta
)\right\rangle $. Helix state $\left\vert \psi _{\mathrm{E}}\right\rangle $\
is an entangled state. This example indicates that if the coefficients $%
\left\{ d_{n}^{\prime }\right\} $ of superposition $\sum_{n}d_{n}^{\prime
}\left\vert \psi _{n}\right\rangle $ are deviated from the set $\left\{
d_{n}\right\} $ a little, the quasi-helix state is probably entangled.

In comparison with the helix states presented in previous work \cite%
{popkov2021phantom,jepsen2022long}, the existence of the set of states $%
\left\{ \left\vert \psi _{n}\right\rangle \right\} $\ are well understood on
the basis of the modified SU(2) symmetry of $H_{0}$. In the presence of $H_{%
\mathrm{I}}$, the SU(2) symmetry is broken, $\left( N+1\right) $-fold
degeneracy is left and the set of states $\left\{ \left\vert \psi
_{n}\right\rangle \right\} $\ are no longer the eigenstates. Nevertheless,
certain appropriately designed external field $H_{\mathrm{I}}$ may provide a
pathway to hybrid the $\left( N+1\right) $-fold degenerate states, forming
the helix state on demand. Similar to the helix state in XXZ chain, the
present helix states contain the information of $H_{0}$, the strength of DM
interaction $D$.

\section{Dynamic generation of helix state}

\label{Dynamic generation of helix state}

In this section, we focus on the preparation of a helix state through a
dynamic way, which is a crucial step in coherent experimental protocol. The
strategy is to take an easily prepared eigenstate of $H_{0}$ as the initial
state, and then add\ $H_{\mathrm{I}}$. It is expected that the evolved state
to be a helix state at certain instant. In the following, we consider three
kinds of $H_{\mathrm{I}}$, which are spatially modulated Hermitian,
non-Hermitian fields, and balanced non-Hermitian boundary respectively.

\subsection{Hermitian field}

We consider the situation that the system is exerted by a resonant field 
\begin{equation}
\mathbf{B}_{j}=B_{0}(t)\left[ \cos \left( k_{0}j\right) ,-\sin \left(
k_{0}j\right) ,0\right] ,
\end{equation}%
where $B_{0}(t)$\ is an arbitrary function of time, but is taken as a pulse
function in our scheme. Here the word resonance does not mean in the
magnitude or frequency but the matching distribution of the field with
coupling strength in the spin chain. We will show that such a spatially
modulated pulse field can drive a simple ferromagnetic state to a precession
helix state.

In general, the time evolution of a given initial state $\left\vert \psi
\left( 0\right) \right\rangle $ under a time dependent Hamiltonian $H(t)$
can be expressed as%
\begin{equation}
\left\vert \psi \left( t\right) \right\rangle =\mathcal{T}\exp \left[
-i\int_{0}^{t}H(t^{\prime })\mathrm{d}t^{\prime }\right] \left\vert \psi
\left( 0\right) \right\rangle ,
\end{equation}%
with $\mathcal{T}$ being the time-ordered operator. The merit of a resonant
field is the commutative relation%
\begin{equation}
\left[ H_{0},H_{\mathrm{I}}\right] =0,
\end{equation}%
which ensures the analytical expression 
\begin{eqnarray}
\left\vert \psi \left( t\right) \right\rangle
&=&e^{it(N-1)/4}e^{-i\int_{0}^{t}H_{\mathrm{I}}(t)\mathrm{d}t}\left\vert
\Downarrow \right\rangle  \notag \\
&=&e^{it(N-1)/4}\overset{N}{\prod\limits_{j=1}}\left\{ -ie^{ik_{0}j}\sin %
\left[ \int_{0}^{t}\frac{1}{2}B_{0}(t^{\prime })\mathrm{d}t^{\prime }\right]
\left\vert \uparrow \right\rangle _{j}\right.  \notag \\
&&\left. +\cos \left[ \int_{0}^{t}\frac{1}{2}B_{0}(t^{\prime })\mathrm{d}%
t^{\prime }\right] \left\vert \downarrow \right\rangle _{j}\right\} .
\end{eqnarray}%
for the initial state $\left\vert \psi \left( 0\right) \right\rangle
=\left\vert \Downarrow \right\rangle $. Obviously, it is a precession helix
state with the vector%
\begin{equation}
\mathbf{h}_{l}(t)=\frac{1}{2}[\sin \theta \sin \left( k_{0}l\right) ,\sin
\theta \cos \left( k_{0}l\right) ,-\cos \theta ],
\end{equation}%
where $\theta $\ is a function of time%
\begin{equation}
\theta (t)=\int_{0}^{t}B_{0}(t^{\prime })\mathrm{d}t^{\prime }.
\end{equation}%
One find that $\left\vert \psi \left( t\right) \right\rangle $ is a helix
state\ at every fixed time point satisfying $\theta =n\pi +\pi /2, \left(
n\in Z\right) $. Specifically, when we take $B_{0}(t)$\ as a pulse field
satisfying $B_{0}(t)=0$\ for $t>T$, and $\int_{0}^{T}B_{0}(t)\mathrm{d}t=\pi
/2$, we have a stable state with maximal helicity

\begin{equation}
\mathbf{h}_{l}(t>T)=\frac{1}{2}[\sin \left( k_{0}l\right) ,\cos \left(
k_{0}l\right) ,0].
\end{equation}

As an example, we consider a Gaussian pulse driving field%
\begin{equation}
B_{0}(t)=\frac{\sqrt{\pi \alpha }}{2}\exp [-\alpha (t-T/2)^{2}],  \label{G}
\end{equation}%
where the internal $T$ is taken sufficiently long as $\alpha \gg T^{-2}$\ to
meet $\int_{0}^{T}B_{0}(t)\mathrm{d}t\approx \pi /2$. Note that the
conclusion is obtained under the resonant condition $\Delta =1$. It is
expected that a similar helix state can still be obtained when $\Delta $
deviates a little from $1$. The computation is performed by using a uniform
mesh in the time discretization for the time-dependent Hamiltonian $H(t)$. 
\begin{figure}[tbh]
\centering
\includegraphics[width=0.5\textwidth]{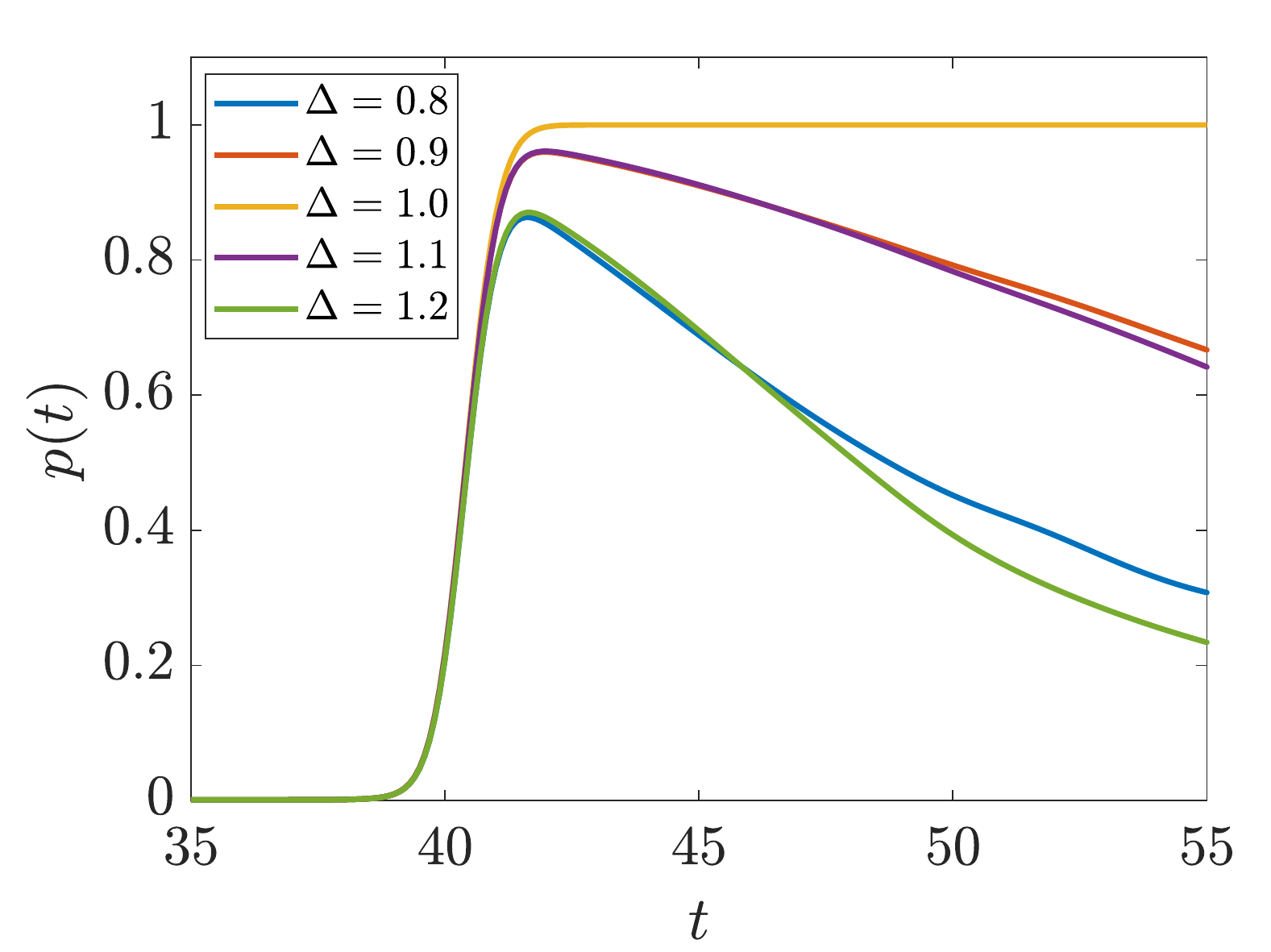}
\caption{Plots of the fidelity defined in Eq. (\protect\ref{fidelity}) for
the time evolution of initial ferromagnetic state under the Hamiltonian $H$
with external field in Eq. (\protect\ref{G}) and different $\Delta $. The
target state is $\left\vert \protect\phi \left( \protect\pi /2\right)
\right\rangle $ and the parameters are $\protect\alpha =0.5$, $T=80$, $%
k_{0}=\arctan (0.5)$ and $N=10$. We find that (i) the fidelity reaches unity
after the action of the pulsed field in the resonant case $\Delta =1$, which
accords with our analytical prediction; (ii) when $\Delta =1\pm 0.1$, the
fidelity reaches a maximum close to unity; (iii) as $\Delta $ departs from $%
1 $, the maximum decreases but still more than $0.8$. The time is in units
of $J^{-1}$ where $J$ is the scale of the Hamiltonian and we take $J=1$.}
\label{fig2}
\end{figure}
We consider the case with initial state $\left\vert \psi \left( 0\right)
\right\rangle =\left\vert \phi \left( 0\right) \right\rangle $. We introduce
the quantity%
\begin{equation}
p\left( t\right) =\left\vert \langle \psi \left( t\right) \left\vert \phi
\left( \theta \right) \right\rangle \right\vert ^{2}  \label{fidelity}
\end{equation}%
to characterize the fidelity of the scheme. The plots of $p\left( t\right) $%
\ in Fig. \ref{fig2} for several typical cases show that the scheme works
well even for the case with $\Delta \neq 1$. However, the flaw of this
scheme is that the prior knowledge of system parameter $k_{0}$ and a
time-dependent field is required.

\subsection{Non-Hermitian field}

Now we turn to alternative scheme to prepare helix state by non-Hermitian $%
H_{\mathrm{I}}$.\ It is a crossover scheme for the case that $k_{0}$ is
unknown. We start with the investigation for an exactly solvable case, in
which the external field is a complex spatially modulated field%
\begin{equation}
\mathbf{B}_{j}=B_{0}e^{ik_{0}j}\left( 1,i,0\right) ,
\end{equation}%
with which we still have $\left[ H_{0},H_{I}\right] =0$. Importantly, we have%
\begin{equation}
H_{I}\left\vert \psi _{n}\right\rangle =B_{0}\sqrt{\left( n+1\right) \left(
N-n\right) }\left\vert \psi _{n+1}\right\rangle ,
\end{equation}%
with $n\in \left[ 0,N-1\right] $, which ensures the existence of an
invariant ($N+1$)-D\ subspace spanned by set of states $\left\{ \left\vert
\psi _{n}\right\rangle \right\} $. The matrix representation of Hamiltonian $%
H$ is an $\left( N+1\right) \times \left( N+1\right) $ matrix $M$ with
nonzero matrix elements

\begin{equation}
\left( M\right) _{N+1-n,N-n}=B_{0}\sqrt{\left( n+1\right) \left( N-n\right) }%
,
\end{equation}%
with $n=\left[ 0,N-1\right] $, and%
\begin{equation}
\left( M\right) _{N+1-n,N+1-n}=-(N-1)/4,
\end{equation}%
with $n=\left[ 0,N\right] $. It is obviously $M+(N-1)/4$ is a nilpotent
matrix, i.e.%
\begin{equation}
\left[ M+(N-1)/4\right] ^{N+1}=0,
\end{equation}%
or an $\left( N+1\right) $-order Jordan block. The dynamics for any states
in this subspace is governed by the time evolution operator%
\begin{equation}
U(t)=e^{-iMt}=\sum_{l=0}^{N}\frac{1}{l!}\left( -iMt\right) ^{l}.
\end{equation}%
Then for the initial state $\left\vert \psi \left( 0\right) \right\rangle
=\left\vert \Downarrow \right\rangle $, we have the normalized evolved state%
\begin{equation}
\left\vert \psi \left( t\right) \right\rangle =\frac{e^{it(N-1)/4}}{\sqrt{%
\left( 1+B_{0}^{2}t^{2}\right) ^{N}}}\overset{N}{\underset{j=1}{\prod }}%
\left( -itB_{0}e^{ik_{0}j}\left\vert \uparrow \right\rangle _{j}+\left\vert
\downarrow \right\rangle _{j}\right) ,
\end{equation}%
which turns to the coalescing state, i.e., $\left\vert \psi \left( \infty
\right) \right\rangle \longrightarrow \left\vert \Uparrow \right\rangle $.
Accordingly, we have%
\begin{equation}
\mathbf{h}_{l}(t)=\frac{B_{0}t}{1+B_{0}^{2}t^{2}}[\sin \left( k_{0}l\right)
,\cos \left( k_{0}l\right) ,\frac{B_{0}^{2}t^{2}-1}{2B_{0}t}],
\end{equation}%
which indicates that $\left\vert \psi \left( t\right) \right\rangle $ is a
helix state at finite time. At instant $t=B_{0}^{-1}$, it reaches the
maximal helicity

\begin{equation}
\mathbf{h}_{l}(B_{0}^{-1})=\frac{1}{2}[\sin \left( k_{0}l\right) ,\cos
\left( k_{0}l\right) ,0].  \label{max h}
\end{equation}%
The above analysis is still true when we take $\mathbf{B}%
_{j}=B_{0}e^{-ik_{0}j}\left( 1,-i,0\right) $\ and $\left\vert \psi \left(
0\right) \right\rangle =\left\vert \Uparrow \right\rangle $, which
corresponds to a time reversal process. 
\begin{figure*}[tbh]
\centering
\includegraphics[width=1\textwidth]{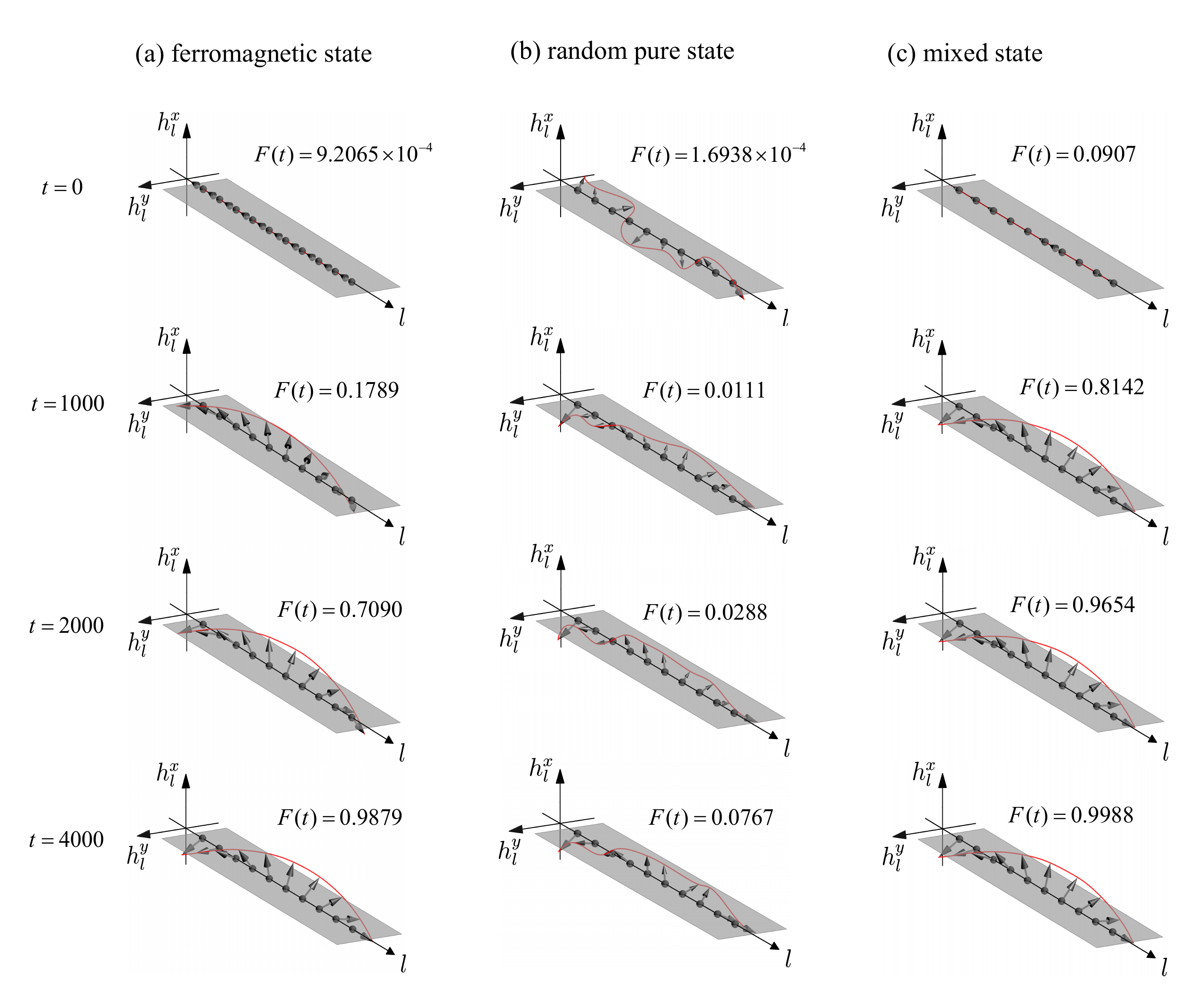}  
\caption{Plots of numerical results of time evolution for three types
initial states under the Hamiltonian $H$ with non-Hermitian boundary in Eq.
( \protect\ref{B}). The initial states are (a) ferromagnetic state, (b) pure
random state and (c) mixed state, which are defined in the text. The
corresponding fidelity $F(t)$ defined in Eq. (\protect\ref{Ft}) is presented
at several typical instants $t$. The complete plot of $F(t)$ is given in
Fig. \protect\ref{fig4}. The parameters are $B_{0}=0.005$, $N=10$ and $%
k_{0}=\arctan (0.5)$. It indicates that the evolved state for initial mixed
state converges faster than that for the other two. The time is in units of $%
J^{-1}$ where $J$ is the scale of the Hamiltonian and we take $J=1$.}
\label{fig3}
\end{figure*}

\subsection{Non-Hermitian boundary}

So far, it seems that the introduction of the complex field does not improve
the scheme since it still requires a specific field distribution. The only
difference is that the time evolution under $U(t)$\ is unidirectional,
rather than periodic in the Hermitian system. However, there is a key fact
that the Jordan block still exists when we take a local complex field at $l$%
th site%
\begin{equation}
\mathbf{B}_{j}=B_{0}\delta _{jl}\left( 1,i,0\right) .
\end{equation}%
Actually, in the case of $\Delta \neq 1$, states $\left\vert \Downarrow
\right\rangle $\ and $\left\vert \Uparrow \right\rangle $\ are two
degenerate states of the Hermitian Hamiltonian $H_{0}$, and we have%
\begin{equation}
H\left\vert \Uparrow \right\rangle =-(N-1)/4\left\vert \Uparrow
\right\rangle ,H^{\dag }\left\vert \Downarrow \right\rangle
=-(N-1)/4\left\vert \Downarrow \right\rangle ,
\end{equation}%
due to the facts%
\begin{equation}
H_{\mathrm{I}}\left\vert \Uparrow \right\rangle =0,\left( H_{\mathrm{I}%
}\right) ^{\dag }\left\vert \Downarrow \right\rangle =0.
\end{equation}%
It means that two states $\left\vert \Downarrow \right\rangle $\ and $%
\left\vert \Uparrow \right\rangle $\ are mutually biorthogonal conjugate and 
$\left\langle \Downarrow \right. \left\vert \Uparrow \right\rangle $ is the
biorthogonal norm of them. Importantly, the vanishing norm $\left\langle
\Downarrow \right. \left\vert \Uparrow \right\rangle =0$ indicates that
state $\left\vert \Uparrow \right\rangle $ ($\left\vert \Downarrow
\right\rangle $) is coalescing state of $H$ ($H^{\dag }$), or Hamiltonians $H
$ and $H^{\dag }$ get an EP. From the perspective of dynamics, we have%
\begin{equation}
e^{-iHt}\left\vert \Downarrow \right\rangle \longrightarrow \left\vert
\Uparrow \right\rangle ,e^{-iH^{\dag }t}\left\vert \Uparrow \right\rangle
\longrightarrow \left\vert \Downarrow \right\rangle ,
\end{equation}%
for a sufficiently long time $t$. Although both states $\left\vert
\Downarrow \right\rangle $\ and $\left\vert \Uparrow \right\rangle $\ are
not helix states, $e^{-iHt}\left\vert \Downarrow \right\rangle \ $and $%
e^{-iH^{\dag }t}\left\vert \Uparrow \right\rangle $ may have helicity at
finite $t$ from the observation at the end of the previous subsection, for
instance, Eq. (\ref{max h}).

This inspires us to consider a balanced local complex field%
\begin{equation}
\mathbf{B}_{j}=B_{0}\left[ \delta _{1j}\left( 1,i,0\right) +\delta
_{Nj}\left( 1,-i,0\right) \right] ,  \label{B}
\end{equation}%
which acts as non-Hermitian boundary and may result in stable helix state
after a relaxation time. The physical intuition for this setup is simple.
One complex field acts as a source of spin flips, while the other one takes
the role of drain. It is expected that a stable helix state emerges when the
source and drain are balanced. However, it is hard to get exact solution in
this case due to the fact $\left[ H_{0},H_{I}\right] \neq 0$. In the
following, we investigate this issue by perturbation method. In the subspace
spanned by the set of degenerate ground states $\left\{ \left\vert \psi
_{n}\right\rangle \right\} $ of $H_{0}$, the matrix representation of
Hamiltonian $H$ with $\Delta =1$ is an $\left( N+1\right) \times \left(
N+1\right) $ matrix $\mathcal{H}$ with nonzero matrix elements

\begin{eqnarray}
\left( \mathcal{H}\right) _{N+1-n,N-n} &=&\frac{B_{0}e^{-ik_{0}}}{N}\sqrt{%
\left( n+1\right) \left( N-n\right) }, \\
\left( \mathcal{H}\right) _{N-n,N+1-n} &=&\frac{B_{0}e^{ik_{0}N}}{N}\sqrt{%
\left( n+1\right) \left( N-n\right) },
\end{eqnarray}%
with $n=\left[ 0,N-1\right] $, and%
\begin{equation}
\left( \mathcal{H}\right) _{N+1-n,N+1-n}=-(N-1)/4,
\end{equation}%
with $n=\left[ 0,N\right] $. In small $B_{0}$\ limit, the eigenvalues and
eigenvectors of matrix $\mathcal{H}$ are the approximate solutions of the
non-Hermitian Hamiltonian. We note that matrix $\mathcal{H}$ is essentially
related to the representation of the Hamiltonian $\mathcal{H}^{^{\prime }}$
of a fictitious spin $S=N/2$ particle: $\mathcal{H}^{^{\prime }}\mathcal{%
=\lambda }S_{x}$, where $S_{x}$ is its angular momentum operator and $%
\mathcal{\lambda }$ is some complex constant. Then the normalized\
approximate eigenstates can be obtained from states $\left\{ \left\vert \psi
_{n}\right\rangle \right\} $%
\begin{equation}
\left\vert \widetilde{\psi }_{n}\right\rangle =R\left\vert \psi
_{n}\right\rangle =\overset{N}{\underset{j=1}{\prod }}R_{j}\left\vert \psi
_{n}\right\rangle ,
\end{equation}%
by a local transformation on spin at each site%
\begin{equation}
R_{j}=\frac{1}{\sqrt{2}}\left( 
\begin{array}{cc}
e^{ik_{0}\left( 2j-N-1\right) /2} & 1 \\ 
1 & -e^{-ik_{0}\left( 2j-N-1\right) /2}%
\end{array}%
\right) .
\end{equation}%
The corresponding eigenenergy is complex%
\begin{equation}
E_{n}=-\frac{N-1}{4}+\frac{B_{0}e^{ik_{0}(N-1)/2}}{N}(2n-N),
\end{equation}%
with $n=\left[ 0,N\right] $ and its imaginary part is%
\begin{equation}
\mathrm{Im}\left( E_{n}\right) =\frac{B_{0}}{N}(2n-N)\sin \left[ \frac{%
k_{0}(N-1)}{2}\right] .
\end{equation}%
Unlike a Hermitian system, the imaginary part of eigenenergy can amplify or
reduce the corresponding amplitude of the eigenstate in the dynamic process.
For the given initial state $\left\vert \psi \left( 0\right) \right\rangle
=\left\vert \Uparrow \right\rangle $, when the evolution time is long enough
the finial state is the eigenstate of $H$ with the maximum imaginary part of
eigenenergy. The corresponding approximate eigenstate is

\begin{equation}
\left\vert \psi \left( \infty \right) \right\rangle =\left\{ 
\begin{array}{cc}
\sum\limits_{n=0}^{N}p_{n}\left\vert \psi _{n}\right\rangle , & \sin \left[ 
\frac{k_{0}(N-1)}{2}\right] >0 \\ 
\sum\limits_{n=0}^{N}(-1)^{n}p_{n}\left\vert \psi _{n}\right\rangle , & \sin 
\left[ \frac{k_{0}(N-1)}{2}\right] <0%
\end{array}%
\right. ,
\end{equation}%
where the coefficient is%
\begin{equation}
p_{n}=2^{-N/2}\sqrt{C_{N}^{n}}e^{-ik_{0}(N+1)n/2}.
\end{equation}%
Accordingly, we have the helicity distribution along the chain%
\begin{equation}
\mathbf{h}_{l}=\frac{1}{2}\left[ \cos \left( k_{0}\ell \right) ,-\sin \left(
k_{0}\ell \right) ,0\right] ,
\end{equation}%
for $\sin \left[ k_{0}(N-1)/2\right] >0$, and 
\begin{equation}
\mathbf{h}_{l}=\frac{1}{2}\left[ -\cos \left( k_{0}\ell \right) ,\sin \left(
k_{0}\ell \right) ,0\right] ,
\end{equation}%
for $\sin \left[ k_{0}(N-1)/2\right] <0$, where $\ell =l-(N+1)/2$\ is a
shifted coordinate. Obviously, the above two classes of state $\left\vert
\psi \left( \infty \right) \right\rangle $ are standard helix states with
opposite helicity, due to the fact $\left\vert \mathbf{h}_{l}\right\vert
^{2}=0.25$.

Numerical simulation is performed to verify our predictions. We compute the
time evolution by exact diagonalization and present the dynamic process of
the formation of the helix state through the time dependence of the helicity
distribution $\mathbf{h}_{l}$. In general, the time evolution of an
arbitrary initial state $\rho \left( 0\right) $ obeys the equation 
\begin{equation}
i\frac{\partial }{\partial t}\rho \left( t\right) =H\rho \left( t\right)
-\rho \left( t\right) H^{\dag },
\end{equation}%
which admits the formal solution 
\begin{equation}
\rho \left( t\right) =e^{-iHt}\rho \left( 0\right) e^{iH^{\dag }t}.
\end{equation}%
Unlike the Hermitian case, the time evolution of the density matrix is no
longer unitary. In order to get $\mathbf{h}_{l}(t)$, with the definition%
\begin{equation}
h_{l}^{\alpha }=\mathrm{Tr}\left[ \rho \left( t\right) s_{l}^{\alpha }\right]
,(\alpha =x,y,z),
\end{equation}%
we normalize $\rho \left( t\right) $ by taking \cite{DCBrody, KKawabata} 
\begin{equation}
\rho \left( t\right) =e^{-i\mathcal{H}t}\rho \left( 0\right) e^{i\mathcal{H}%
^{\dag }t}/\mathrm{Tr}\left[ e^{-i\mathcal{H}t}\rho \left( 0\right) e^{i%
\mathcal{H}^{\dag }t}\right] ,
\end{equation}%
in the following numerical calculation. We introduce the Uhlmann fidelity 
\cite{AUhlmann, NTJacobson} 
\begin{equation}
F\left( t\right) =\left[ \mathrm{Tr}\sqrt{\sqrt{\rho _{\mathrm{h}}}\rho
\left( t\right) \sqrt{\rho _{\mathrm{h}}}}\right] ^{2},  \label{Ft}
\end{equation}%
to characterize the degree of similarity between the evolved state $\rho
\left( t\right) $ and the target state 
\begin{equation}
\rho _{\mathrm{h}}=\left\vert \psi \left( \infty \right) \right\rangle
\left\langle \psi \left( \infty \right) \right\vert .
\end{equation}%
The value of $F\left( t\right) $ after a sufficient long time can be
estimated intuitively. In general, an initial mixed state $\rho \left(
0\right) $ contains equal-amplitude components in each state of $\left\{
\left\vert \psi _{n}\right\rangle \right\} $. Then we always have $F\left(
\infty \right) \approx 1$.

We focus on three types of initial states: (i) ferromagnetic state $\rho
\left( 0\right) =\left\vert \Uparrow \right\rangle \left\langle \Uparrow
\right\vert $; (ii) random pure state $\rho \left( 0\right) =\left\vert \psi
\left( 0\right) \right\rangle \left\langle \psi \left( 0\right) \right\vert $%
, where 
\begin{equation}
\left\vert \psi \left( 0\right) \right\rangle =\left[ \sum_{n=1}^{2^{N}}%
\left( \alpha _{n}\right) ^{2}\right] ^{-1/2}\sum_{n=1}^{2^{N}}\alpha
_{n}\left\vert n\right\rangle .
\end{equation}%
Here coefficient $\alpha _{n} $ is taken as a uniform random number within
the interval $\left( -1,1\right) $, and $\left\{ \left\vert n\right\rangle
\right\} $ is the complete set of eigenstates of $H_{0}$; (iii) maximally
mixed ferromagnetic state 
\begin{equation}
\rho \left( 0\right) =\frac{1}{N+1}\sum_{n=0}^{N}\left\vert \psi
_{n}\right\rangle \left\langle \psi _{n}\right\vert .
\end{equation}

\begin{figure}[tbh]
\centering
\includegraphics[width=0.5\textwidth]{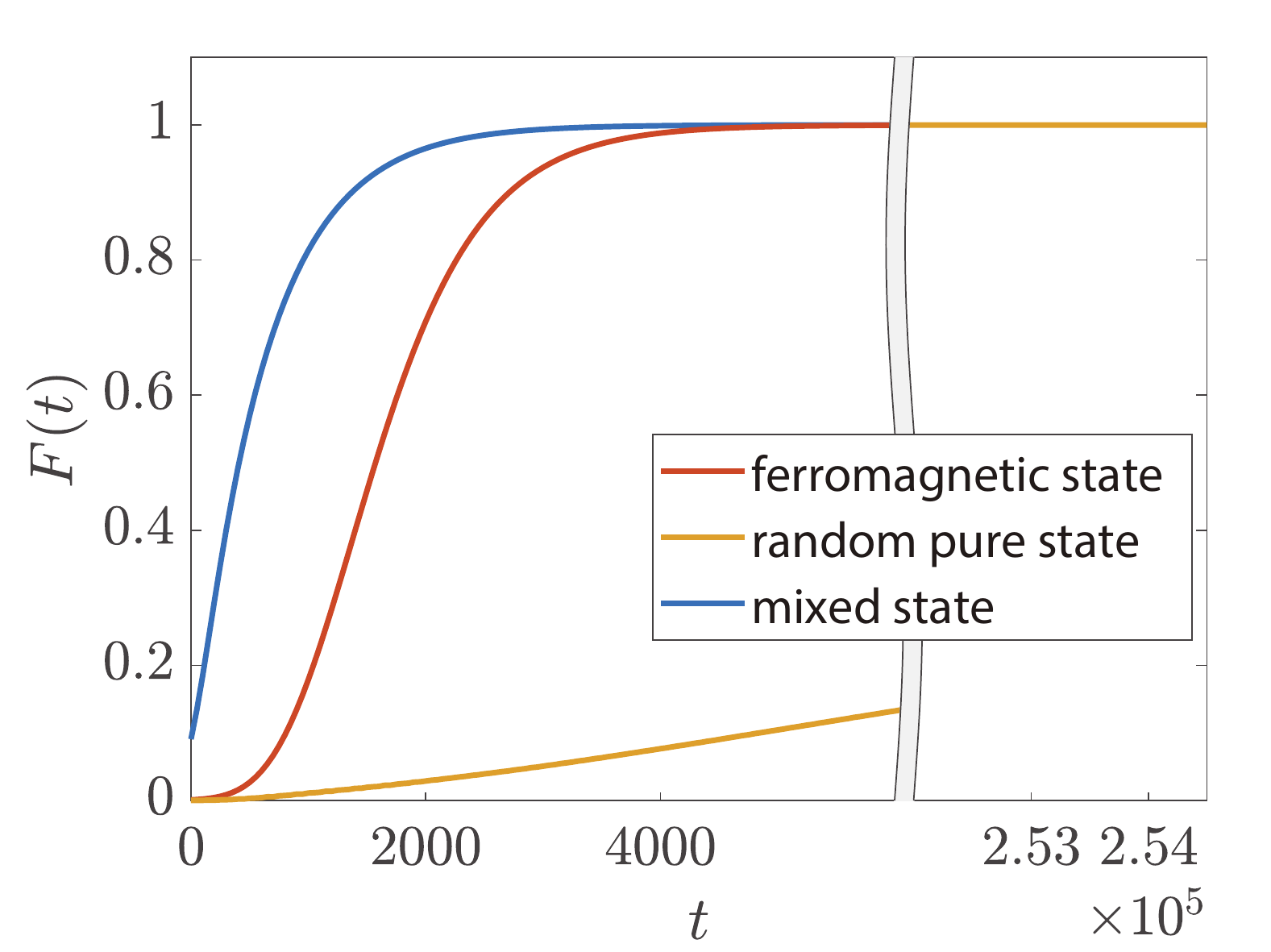}  
\caption{Plots of $F(t)$ in Eq. (\protect\ref{Ft}) as a function of time for
same time evolution process in Fig. \protect\ref{fig3}. We can see that the
final states for three different initial states turn to the target state
eventually. The time is in units of $J^{-1}$ where $J$ is the scale of the
Hamiltonian and we take $J=1$.}
\label{fig4}
\end{figure}

The plots of $\mathbf{h}_{l}$ and $F\left( t\right) $ in Figs. \ref{fig3}
and \ref{fig4} show the dynamic behaviors of the evolved states of above
three types of initial states, induced by the non-Hermitian boundary. It
indicates that the evolved states for initial mixed state and ferromagnetic
state converge fastly. Importantly, the final states for all three different
initial states turn to the target state after sufficient long time. Notably,
the initial states, as well as the selected non-Hermitian boundary, do not
contain any information of the prequench Hamiltonian. 

\section{Summary}

\label{Summary}

In summary, we have studied the possible helix states in XXZ Heisenberg
model with DM interaction. Unlike the previous works on this topic, the
existence of spin helix state in this work is the direct result of the
resonant DM interaction. Our findings offer a method for the efficient
preparation of a spin helix state as the ground state of a spin chain by the
quench dynamic process with the aid of non-Hermitian balanced perturbation.
It is expected to be insightful for quantum engineering by non-Hermitian
boundary.

\acknowledgments This work was supported by National Natural Science
Foundation of China (under Grant No. 11874225).

\end{document}